\documentclass[12pt]{article}
\usepackage{latexsym}

\def\tr{{\rm tr}}
\def\ket#1{\mid~\!\!\!{#1}~\!\!\rangle}
\def\bra#1{\langle~\!\!{#1}~\!\!\!\mid}

\def\QM{quantum mechanics }
\def\qm{quantum mechanics}
\def\Q{quantum }

\def\QMl{quantum-mechanical }
\def\qml{quantum-mechanical}
\def\${\enskip$}
\def\M{measurement }
\def\m{measurement}
\def\I{interpretation }

\begin{document}

\begin{center}
{\bf A Theory of Quantum Preparation}\\
\vspace{0.2cm}

F Herbut\\
\vspace{0.2cm}

Serbian Academy of
Sciences and Arts, Knez Mihajlova
35, 11000 Belgrade, Serbia\\
\vspace{0.2cm}

E-mail: fedorh@sanu.ac.rs\\
\vspace{0.2cm}

\date{\today}
\vspace{0.2cm}

\end{center}

\begin{abstract}\noindent
Based on an analysis of two
conventional preparators, the
Stern-Gerlach and the
hole-in-the-screen ones, it is
argued that four entities can be taken as the
basic ingredients of a rather general theory of a
quantum preparator. These are the
composite-system (object plus preparator) state coming about as a
result of a suitable interaction
between the subsystems, a suitable
preparator projector called the
triggering event, the conditional
quantum state (density operator)
of the quantum object coming about
as a consequence of the occurrence
of the triggering event on the preparator, and, finally, a unitary evolution
operator of the object subsystem
acting after preparation. The
concepts of a general conditional state
and of retrospective apparent
ideal occurrence (which appears in
the theory) are discussed in considerable
detail. Ideal occurrence and the selective L\"{u}ders formula, which are made use of, are reviewed. Dynamical and geometrical
preparators are distinguished in
the general theory. They are
described by the same entities in
the same way, but in terms of
different physical mechanisms from the point of view of standard interpretation with collapse.\\
\end{abstract}

PACS numbers: 03.65.-w, 03.65.Ca,
03.65.Ta

Key words: Experiment, conditional
state, retrospective apparent
ideal occurrence, puzzles

 \normalsize \rm

%Intr
%%%%%%%%%%%%%%%%%%%%%%%%%%%%
\section{Introduction}
%%%%%%%%%%%%%%%%%%%%%%%%%%%%

Quantum physics is concerned with
quantum experiments. These have
three parts: preparation, (time)
evolution, and \m . The {\it first motivation}
for this article is to present a \QMl
understanding of the first part in a class of experiments because it is by far less elaborated in the literature than the \M at the end of the experiment. The {\it second motivation} is an attempt to find out why has preparation received so little attention in the literature.

It will turn out that there are two kinds of preparations: dynamical and geometrical ones. It is intended to treat both kinds
by the same formalism in order to keep preparation theory as simple as possible. {\it This goal} is achieved at the expense of introducing a fictitious event in preparation. It is fictitious as far as the standard quantum-mechanical concept of occurrence of an event is concerned. But it will be seen, as a rule, to be quite real concerning classical intuition.

In the present study a central role is played by an event (projector) \$Q\$ on the preparator, called 'triggering' event. Hence, it must be clarified what {\it occurrence} of an event means in the standard Bohrian Copenhagen interpretation that pervades the textbooks of \qm . Let us be reminded of a known definition by Bohr himself \cite{Bohr}

\begin{quote}
"As a more appropriate way of expression I
advocate the application of the word
{\it phenomenon} exclusively to refer
to the observations obtained under
specified circumstances, including an
account of the whole experimental
arrangement.(The italics are Bohr's.)
\end{quote}

I think that Bohr means by "phenomenon"
a {\it real phenomenon}, i. e., that
this is where 'reality', i. e., real occurrence, enters the
scene in the view of Bohr. In the textbooks one says that an event takes place in \QM if an event is measured by some kind of classical experimental arrangement.

\vspace{0.5cm}

%SG sec
%%%%%%%%%%%%%%%%%%%%%%%%%%%%%%%%%%%%%%%%
\section{Stern-Gerlach Preparators}
%%%%%%%%%%%%%%%%%%%%%%%%%%%%%%%%%%%%%%%%

To begin with, let us sum up some
of the familiar  aspects of
Stern-Gerlach (SG)
spin-projection \M in order to
single out those of its features
that are relevant for a theory of
a quantum preparator.\\

%SG M subsec
%%%%%%%%%%%%%%%%%%%%%%%%%%%%%%%%%%%%%
\subsection{Stern-Gerlach \M }
%%%%%%%%%%%%%%%%%%%%%%%%%%%%%%%%%%%%%

We assume that it is the
z-projection of the spin of a
spin-one-half particle that is
measured. The incoming particle is
in the uncorrelated pure state
given by a state vector of the form
$$\ket{\Psi}_{I,II}\equiv\bigg(\alpha
\ket{+,z}_I+\beta\ket{-,z}_I
\bigg)\ket{\psi^{in}} _{II}.$$ Here
\$\alpha ,\beta \in \mbox{\bf C},
\enskip |\alpha |^2 +
|\beta|^2=1\$. The first subsystem
is that of spin one-half and the
second one consists of the spatial
degrees of freedom of the
particle. (Note that \qml ly
subsystems need not be material
ones; they can be degrees of
freedom as in this case.) Further,
\$\ket{\pm ,z}_I\$ are the spin-up
and spin-down (along z) state
vectors. Finally,
\$\ket{\psi^{in
}_{II}}\$ is the
incoming spatial state vector of
the particle.

As well known, the magnetic field
couples the z-projection of spin
with the spatial degrees of
freedom of the outgoing particle
(leaving the field and reaching
the plates) as follows. Let us
denote by \$\ket{\psi^+}_{II}\$
the upward moving particle state
(which reaches the upper plate), and
by \$\ket{\psi^-} _{II}\$ the
downward moving particle state
(reaching the lower plate). Then
$$ \ket{\Phi}_{I,II}\equiv\alpha
\ket{+,z}_I\ket{\psi ^+}_{II} +
\beta \ket{-,z}_I\ket{\psi
^-}_{II} \eqno{(1)}$$ is the
composite-system state after the
coupling.

Let us introduce the projectors
$$ Q^+_{II}\equiv\int^{+\infty}_
{-\infty}
\int^{+\infty}_{-\infty}\int
^{+\infty}_0\ket{x,y,z}_{II}\bra{
x,y,z}_{II}dxdydz,\eqno{(2a)}$$
$$ Q^-_{II}\equiv
\int^{+\infty}_ {-\infty}
\int^{+\infty}_{-\infty}\int
^0_{-\infty }\ket{x,y,z}_{II}\bra{
x,y,z}_{II}dxdydz\eqno{(2b)}$$
projecting onto the upper and the
lower half-spaces respectively.

One should note that
$$Q_{II}^-=I_{II}-Q_{II}^+\equiv
(Q_{II}^+)^{\perp},$$ \$I_{II}\$
being the identity operator for
the spatial subsystem. Thus,
\$Q_{II}^-\$ is determined by (2a).

Ideal occurrence of the
second-subsystem event
\$(I_I\otimes Q^+_{II})\$ (usually
written simply as \$Q^+_{II}\$) in
the composite-system state
\$\ket{\Phi}_{I,II}\$ given by (1)
brings the entire system, as it is
known, into the state
$$cQ^+_{II}\ket{\Phi}_{I,II}=
\ket{+,z}_I\ket{\psi ^+}_{II},
$$ where \$c\$ is the
corresponding (positive)
normalization constant. Namely,
according to the selective
L\"{u}ders formula \cite{Lud},
\cite{Messiah}, \cite{Laloe}, one
applies the projector onto the
state vector and one renormalizes
the result. By this, the spin
subsystem is brought into the
state $$\ket{+,z}_I.\eqno{(3)}$$

The SG measuring apparatus in its
standard form performs
retrospective or second-kind
measurement when the particle is
stopped on one of the plates.
This causes drastic change in the
spatial state of the particle
(spot on the plate) and a
corresponding change in the state
of the measuring instrument.\\

In spite of the above usual simple
theory, the SG \M is actually
so-called distant \M
\cite{FHMV'76}: the spin
projections and 'being in the
upper or lower half-space' are
so-called twin observables in (1).
Interaction takes place only
between the measuring instrument
and the spatial degrees of freedom
(subsystem \$II\$) of the particle
leading to direct \M of the
latter; but this is then, by this
very act, an interaction-free \m ,
called distant \m , of the spin
projection on account of the
strong correlations that give rise
to the twin observables in (1). We
do not need to go into the
niceties of this conceptually
intricate measuring process. (More
details can be found in
\cite{FHMV'76}, \cite{FHMV'84}, as
well as in \cite{FHPhys.Rev.} and
in the
references in it.)\\

To obtain a preparator,
modification is required.\\

%SG prep 1 subsec
%%%%%%%%%%%%%%%%%%%%%%%%%%%%%%%%%%%%%%%
\subsection{The first modification
for a SG preparator}%%%%%%%%%%%%%%%%%%

Let us imagine the following modification of the SG measuring arrangement. The upper plate is replaced by a detector that detects the arrival of the particle via spatial interaction only; and it lets the particle continue its movement to the right out of the arrangement. The point is that the \M affects only the spatial degree of freedom of the particle, and not at all its spin projection. Therefore, it can be considered a spin-projection \m , in which the spin-up event occurs.

In the described \M the particle is detected at the place of the upper plate,
and the purely spatial interaction changes the corresponding component state \$\ket{\psi ^+}_{II}\$ in (1) into an outgoing spatial state with the spin state unchanged: $$\ket{\Phi}_{I,II}\quad
\rightarrow\quad\ket{+,z}_I \ket{\psi ^{out}}_{II}.\eqno{(4)}$$

One can evaluate the spin state as a conditional state from the composite state (1) with the condition \$Q_{II}^+\$ (cf (2a)) by {\it the general conditional-state formula} $$\rho_I\equiv \tr_{II}\Big(Q_{II}^+\rho_{I,II}\Big)\Big/
\tr\Big(Q_{II}^+\rho_{I,II}\Big)=
\tr_{II}\Big(Q_{II}^+\ket{\Phi}_{I,II} \bra{\Phi}_{I,II}\Big)\Big/(1/2), \eqno{(5)}$$ where \$\tr_{II}\$ denotes the partial trace over the second subsystem (the spatial tensor factor space in this case), and the total trace (\$\tr_{I,II}\$
in this case) is written throughout without subscripts. Substituting the bipartite state vector from (1) then gives in a straightforward manner
$$\rho_I=\ket{+,z}_I\bra{+,z}_I$$ in full agreement with (4).

The spin projection further does not change until it is measured at \$t_f\$, the final moment of the experiment.

One is again dealing with distant \M
in this preparation. (More about distant \M is given in section 6.)\\

%SG prep 2 subsec
%%%%%%%%%%%%%%%%%%%%%%%%%%%%%%%%%%%%%%%
\subsection{The second modification
for a SG preparator}%%%%%%%%%%%%%%%%%%

The upper plate is removed. The
particle that would hit the upper
plate in the standard SG
instrument may now freely leave
the instrument. The lower plate is
in place or, perhaps, it is also
removed and replaced by some other
(more suitable) particle detector.

We want a so-called {\it negative
measurement}: it consists in
the anti-coincidence of arrival of
the particle on the plates and of
non-detection on the lower plate.
This amounts to {\it ideal
occurrence} of 'arrival of the particle in the upper half space', i. e., the occurrence of the event \$Q_{II}^+\$ (cf (2a)).

One should note that \$Q_{II}^+\$  really happens (in contrast to fictitious events introduced below). The conditional-state evaluation (5) is again applicable. (More about negative \M can be found in section 6.)

The described anti-coincidence is
hard to achieve in  the laboratory
because it is not easy to make
certain when the particle is
supposed to arrive at the plates.
Nevertheless it is possible in
principle. Next, we design a more
realistic preparator.\\

%SG prep subsec
%%%%%%%%%%%%%%%%%%%%%%%%%%%%%%%%%%%%%%%
\subsection{The third modification for a
SG preparator}%%%%%%%%%%%%%%%%%%%%%%%%

We remove the upper plate, and leave the lower one in place.
The {\it geometry} is such that it makes
it possible to confine our
interest to the upper half-space,
where, further to the right, we put the measuring apparatus. If it measures anything on the particle (at the final
moment $t_f$ of the experiment),
and one obtains a result, then,
due to the geometry, the particle arriving at the detector
must be in the upper half-space.
Therefore, it must be in the state
$$\ket{+,z}_I\bigg(U_{II}(t_f-t_i,t_i)
\ket{\psi^+}_{II}\bigg),
$$ where \$U_{II}(...)\$ is a unitary
evolution operator in the state
space of the second (spatial)
subsystem (the spin does not
change), and \$t_i\$ is throughout this study the
initial moment of the experiment
(and the final moment of preparation if it is not instantaneous).

This
amounts to the same as if we had
occurrence of the event
\$ Q_{II}^+\$ (cf (2a)) at
\$t_i\$ in the state
\$\ket{\Phi}_{I,II}\$ (and
subsequent evolution).
This fictitious occurrence will be explained in detail in section 5.\\

We have obtained {\it three basic
entities}  for a {\it preparator
theory}:
$$\ket{\Phi}_{I,II}\quad
\qquad Q_{II}^+\quad\quad
\qquad\ket{+,z}_I
\eqno{(6a,b,c)}$$ (cf (1), (2a) and
(3) respectively). We call
\$Q_{II}^+\$ \it the triggering
event \rm as the event the
occurrence of which brings about
the (conditional) prepared subsystem state
\$\ket{+,z}_I\$ (via (5)).

The SG composite-system state
\$\ket{\Phi}_{I,II}\$ given by (1)
is very simple. In particular, the
action of the projector
\$Q_{II}^+\$ on this state leaves
the object in the pure subsystem
state \$\ket{+,z}_I\$. To
check
 if (6a,b,c) are the relevant
entities for preparation also in other cases,
we take another, a quite different
and very well known example.\\

\vspace{0.5cm}

%Hole Prep sec
%%%%%%%%%%%%%%%%%%%%%%%%%%%%%%%%%%%%%%%
\section{Preparation
Through a Hole in the Screen}
%%%%%%%%%%%%%%%%%%%%%%%%%%%%%%%%%%%%%%%

In one-hole preparation the first
subsystem is the particle, the
second is the screen. We think of
the screen as of an infinite
surface perpendicular to the
motion of the incoming particle.

The screen is thought of as broken
up into two non-overlapping
segments: the hole is one of them
(segment \$h\$) and the rest of the
screen is the other (segment
\$rs\$). Hitting the latter, i.e.,
transfer of linear momentum at
this segment, corresponds to the
occurrence of, say, the event (projector)
\$Q^{rs}_{II}\$.

Let us think of a {\it negative \M } consisting in the arrival of the particle at the screen and the non-occurrence of \$Q^{rs}_{II}\$. Then the particle passes the hole, and one has {\it ideal occurrence} of the opposite event
$$Q^h_{II}=I_{II}-Q^{rs}_{II}\eqno{(7)}$$ on the screen.

For simplicity, we assume that the
composite-system at the end of the preparation interaction is in a pure
state \$\ket{\Phi ,t_i}_{I,II}\$.
The event \$Q_{II}^h\$ is {\it the
triggering event}. The state
of the particle when the
preparation is completed, i. e.,
{\it the prepared state}, is due to the {\it ideal occurrence} of the triggering event. It is given by taking the reduced density operator (the subsystem state) after projection and renormalization (the selective L\"{u}ders formula in the pure-state case):
$$\rho_I^h(t_i)\equiv\tr_{II}
\bigg[\bigg(c'Q_{II}^h\ket{\Phi
,t_i}_{I,II}\bigg)\bigg(c'\bra{\Phi
,t_i}_{I,II}Q_{II}^h\bigg)\bigg],
\eqno{(8a)}$$ where \$c'\$ is the (positive)
normalization constant. It is, of
course, assumed that \$Q_{II}^h
\ket{\Phi ,t_i}_{I,II}\not= 0\$,
i.e., that the process considered
allows passage through the hole
with positive probability. The
reduced density operator given by
(8a) is more often written in the
simpler and more explicit form:
$$\rho_I^h(t_i)=\tr_{II}
\Big((\ket{\Phi
,t_i}_{I,II}\bra{\Phi
,t_i}_{I,II})Q_{II}^h\Big)\Big/
\Big[\tr\Big((\ket{\Phi
,t_i}_{I,II}\bra{\Phi
,t_i}_{I,II})Q_{II}^h\Big)\Big].
\eqno{(8b)}$$ (This is possible
due to the commutation of the
operator \$Q_{II}^h\$ with the
other operator \$\ket{\Phi
,t_i}_{I,II}\bra{\Phi
,t_i}_{I,II}Q_{II}^h\$ under the
partial trace in (8a), and due to
idempotency of the former
operator. It is easy to prove
that the subsystem operator
\$Q_{II}^h\$ has the stated
commutation property under the
partial trace over the same
subsystem \$II\$ just like it is
usual under a full trace.)

One can see by comparing (8b) with (5) that the prepared state
\$\rho_I^h(t_i)\$ in this case is
determined by the composite-system
state and the triggering event
{\it in the same way} as in the
case of the SG device.

In analogy with
the modifications
of the SG device (in the preceding section), we can think of
modifications of the hole-preparator. We give just one of them.\\

%Hole Prep mod
Let us take a {\it geometrical} modification, in which no real occurrence of event takes place in the preparation. The geometry is such that if anything is
measured on the particle to the
right of the screen at
\$t_f\$, the
former must have passed the hole,
i.e., it is as if the triggering
event had occurred at \$t_i\$.
(This will be discussed in detail
in section 5.)\\

We have now a good deal of
inductive insight for a  general
quantum mechanical theory of
preparation. Nevertheless, it is
desirable to clarify several
important points
first.\\

\vspace{0.5cm}

%Cond State sec
%%%%%%%%%%%%%%%%%%%%%%%%%%%%%%%%%%%%%%
\section{General Conditional State} %%%%%%%%%%%%%%%%%%%%%%%%%%%%%%%%%%%%%%

Let \$\rho_{I,II}\$ be an
arbitrary given composite-system
(mixed or pure) state (a density
operator). Let, further,
\$Q_{II}\$ be a second-subsystem
event (projector) and let it {\it
occur in whatever way} in the
state \$\rho_{I,II}\$. We want an
answer to the question: In what
state \$\bar\rho_I\$ leaves this
occurrence the first subsystem? We
had an answer for ideal occurrence
via the L\"{u}ders formula in the
preceding subsection (cf (8b)).
Now we are interested in the
answer in the case of a general
occurrence of some event
\$Q_{II}$. The answer is known,
but not well known.

The sought-for state (density
operator) \$\bar\rho_I\$ gives
probability prediction for an
arbitrary first-subsystem event
(projector) \$P_I\$ through the
\QMl probability formula in the trace-rule form
\$\tr\Big(\bar\rho_IP_I\Big)\$,
and, as it is known from the theorem of Gleason \cite{Gleason}, \$\bar\rho_I\$
is, in its turn, {\it determined by the
totality} of all possible
projectors \$P_I$ via this same
formula.

Since an {\it arbitrary} first-subsystem event \$P_I\$ and the given event \$Q_{II}\$ are
compatible events (commuting
projectors), their coincidence on
the one hand and the occurrence of
\$P_I\$ \it immediately after \rm
that of \$Q_{II}\$ on the other
can be considered as one and the
same thing. The coincidence
probability can be written in a factorized form:
%%%%%%
$$\tr\Big(\rho_{I,II}(P_I\otimes Q_{II})
\Big)=\tr_I\Big[\Big(
\tr_{II}(\rho_{I,II}Q_{II})\Big)
P_I\Big]=\Big(\tr(\rho_{I,II}Q_{II})\Big)
\Big(\tr(\bar\rho_IP_I
)\Big),\eqno{(9)}$$
%%%%%
where
%%%%
$$\bar\rho_I\equiv\tr_{II}
(\rho_{I,II}Q_{II})
\Big/\Big(\tr(
\rho_{I,II}Q_{II})\Big),\eqno{(10)}$$
%%%%
and
\$\tr\Big(\rho_{I,II}Q_{II}\Big)\$
is the probability of the event
\$Q_{II}\$ in \$\rho_{I,II}\$.
(Note that in the first equation
in (9) use is made of the fact
that \$P_I\$ behaves under the
opposite partial trace
\$\tr_{II}\$ as a constant does under a
full trace, i. e., it can be taken
outside the partial trace. This is
easily proved. Further, it is easily seen that \$\bar\rho_I\$ given by (10) is a density operator.)

Coincidence can be thought of as
taking place in one measurement,
hence (9) can be viewed
classically, as the well-known conditional-probability formula.
In particular, the second factor
\$\tr(\bar\rho_IP_I)\$
in the third expression in (9) is
then, by definition, the \it
conditional probability \rm of
\$P_I\$ in the state
\$\bar\rho_I\$ in which the second
subsystem is left (immediately)
after \$Q_{II}\$ (the condition) has occurred in \$\rho _{I,II}$.

One should note that since \$P_I\$ is an
arbitrary event, one has \$\bar\rho_I\$ given by (10) is the sought-for
expression for the {\it
conditional state}. Thus, (10) {\it extends} the
partial-trace evaluation in (8b)
to the {\it general case of occurrence}
of \$Q_{II}\$ as a condition.\\

Now we can conclude, without
discussion of more intricate
examples of preparation, that we
can abandon the above restrictions
to pure states and ideal
occurrence of the triggering
event. In a {\it general theory of
preparation} we have so far the following
three crucial entities: the
composite-system (object plus
preparator) state \$\rho_{I,II}(t_i)\$
at the initial instant \$t_i\$ of
the experiment, a triggering event
\$Q_{II}\$, and, finally, the (conditional)
prepared state \$\rho_I(t_i)\$, which is
determined by the preceding two
entities (cf (10)).\\

%RAIO sec
%%%%%%%%%%%%%%%%%%%%%%%%%%%%%%%%%%%%%
\section{Retroactive Apparent Ideal Occurrence}
%%%%%%%%%%%%%%%%%%%%%%%%%%%%%%%%%%%%%

When there is no detector in the
preparator, i.e., when  it is no
measurement at all (the third modification in the SG case and the modification in the hole preparator discussed above),
the prepared state \$\rho
_I(t_i)\$ given by (8b), e. g., has, nevertheless, the
meaning of a conditional state,
assuming validity under the
fictitious condition that a
triggering event \$Q_{II}\$ occurs
in the composite-system state
\$\rho_{I,II}\$ at \$t_i$.

In the cases at issue there is no
actual occurrence of any event
until \$t_f\$, when a measurement
result is obtained. Then, owing to
the {\it geometry} of the experiment,
this amounts to the same, as it
was claimed above, as if
\$Q_{II}\$ had occurred in
\$\rho_{I,II}\$. More precise explanation is desirable.

Every measuring arrangement is located in some (spatial) region \$\bf R\$,
and if a detection event, e. g. \$F\$, occurs in any kind of \M
with a positive probability, then one has \$F\leq
P(\mbox{\bf R})\enskip \Big[\equiv\Big(F=FP(\mbox{\bf R})\Big)
\Big]\$ meaning physically that, by
{\it implication}, also \$P(\mbox{\bf R})\$ occurs, which, in turn, means physically that the measured
quantum system is found in the region \$\mbox{\bf R}\$.\\

{\bf Lemma on Localization} If the mentioned implication of events is valid, then
$$\tr(F\rho )=\Big[\tr\Big(P(\mbox{\bf R})\rho\Big)\Big]
\Big[\tr\Big(F\rho'\Big)\Big],$$ where
$$\rho'\equiv P(\mbox{\bf R})\rho P(\mbox{\bf R})\Big/\tr\Big(P(\mbox{\bf R})\rho\Big).$$

In words, the probability of an event localized in {\bf R}  equals
the product of the probability of localization and
the probability of the same event in the state \$\rho'\$, which is the collapsed state (evaluated via the selective L\"{u}ders formula) due to the occurrence of the localization event \$P(\mbox{\bf R})\$ in an ideal way.

{\bf Proof} Utilizing the above implication, the projector
idempotency and commutation under the trace, one has
$$\tr(F\rho )=\tr\Big(P(\mbox{\bf R}))F\rho \Big)=\tr\Big(FP(\mbox{\bf R})
\rho P(\mbox{\bf R})\Big),$$ and from this the claimed relation immediately follows.\hfill $\Box$\\

We have in mind an experiment, in which an initial state \$\rho(t_i)\$ evolves, assuming the system is dynamically isolated from its environment
in the interval from \$t_i\$ till \$t_f\$, unitarily into the final state \$\rho(t_f)\$.\\

Now we introduce the notion of retroactive apparent ideal occurrence (RAIO).

{\bf Theorem on RAIO}
Let \$Q\$ and \$P\$ be two events (projectors) satisfying
the following conditions:

(i) \$0<\tr\Big(Q\rho(t_i)\Big)<1
\$, i. e., both \$Q\$ and the opposite event \$Q^{\perp}\enskip\Big(\equiv (1-Q)
\Big)\$ can occur, i. e., have a positive probability, in \$\rho(t_i)\$,

(ii) the occurrence of
\$Q\$ in \$\rho(t_i)\$ in an
ideal way makes \$P\$ certain in \$\rho(t_f)\$ , and also dually,

(iii) the occurrence of \$Q^{\perp}\$ in \$\rho(t_i)\$
in an ideal way makes \$P^{\perp}\$ certain in \$\rho(t_f)
\$.

Then the following relation is satisfied. $$P\rho(t_f)P
/\Big[\tr\Big(P\rho(t_f)\Big)\Big]=
U\Big\{Q\rho(t_i)Q/\Big[\tr\Big(Q\rho(t_i)\Big)\Big]
\Big\}U^{\dag}.\eqno{(11)}$$

In words, we obtain the same state if, on the one hand,
the system evolves from \$t_i\$ till \$t_f\$ and then
\$P\$ occurs in an ideal way, and, on the other hand,
when \$Q\$ occurs in an ideal way in \$\rho(t_i)\$ and
then the system evolves till \$t_f\$.

Proof is given in \cite{RAIO2}. (The cited previous article of the present author is devoted to the, somewhat intricate, proof of this theorem, and to its applications among which also preparation is mentioned.)\\

If any measurement result is
obtained {\it in whatever \M } on the particle at
\$t_f\$, this takes place in a
certain spatial region {\bf R} , e. g., to the right of the screen in the hole-preparator
example if the particle approaches
the screen before \$t_i\$ from the
left. Hence, the mentioned result
of \M implies the occurrence of
the event \$P_I(\mbox{\bf R})\$, by which it
is meant that the particle is found in
the mentioned region {\bf R} .

If the triggering event \$Q_{II}\$
does occur in the composite-system
state \$\rho_{I,II}(t_i)\$, then
the event \$P_I(\mbox{\bf R})\$ is {\it
certain} to occur in the state
\$\rho(t_f)\$, the
final moment of the experiment.
This means that the particle must
reach the region {\bf R} . Moreover,
if the triggering event does not
occur, i. e., if the opposite
event \$Q_{II}^{\perp}\$ occurs, at
\$t_i\$ in \$\rho_{I,II}\$, then
\$P_I(\mbox{\bf R})\$ will not occur, i.e.,
\$[I_I-P_I(\mbox{\bf R})]\$ will occur in \$\rho(t_f)\$ - the particle does not reach region {\bf R} .\\

Let us resort now to the special application of the, still general, formula (11).

The above Theorem on RAIO implies:
$$\rho_{I,II}(t_f)^{\mbox{\bf R}}=P_I(\mbox{\bf R})
\Big[U_{I,II}\rho_{I,II}
(t_i)U^{\dagger}_{I,II}\Big]P_I(\mbox{\bf R})\Big/
\Big[\tr\Big(P_I(\mbox{\bf R})
U_{I,II}\rho_{I,II}(t_i)U^{\dagger
}_{I,II}\Big)\Big]=$$
$$U_{I,II}\Big\{Q_{II}\rho_{I,II}(t_i)
Q_{II}\Big/\Big[\tr\Big(
Q_{II}\rho_{I,II}(t_i)\Big)\Big]\Big\}
U^{\dagger }_{I,II}.\eqno{(12)}$$

This means that one would obtain {\it
the same} localized final state \$\rho_{I,II}(t_f)^{\bf R}\$ if, on the
one hand, the event \$P_I(\mbox{\bf R})\$
occurred ideally at \$t_f\$ in the final
state \$\rho_{I,II}(t_f)\$, i. e., if we restricted the final state to the spatial region {\bf R} ,
and, on the other
hand, if the triggering event
\$Q_{II}\$ would occur ideally at
\$t_i\$ in the initial state
\$\rho_{I,II}(t_i)\$ (which actually evolves into \$\rho_{I,II}(t_f)\$), and then the
system evolved in the collapsed state till \$t_f\$. Naturally, the composite system system being at \$t_f\$, actually,  in the state \$\rho_{I,II}(t_f)\$, it is found with probability \$\tr\Big(\rho_{I,II}(t_f)P(\mbox{\bf R})\Big)\$ in the spatial region {\bf R} .

If one utilizes the rhs of (12)
instead of its lhs, then one says
that one has \it retroactive
apparent ideal occurrence \rm
(RAIO) of the event \$Q_{II}\$
in \$\rho_{I,II}(t_i)\$. This is, according to (12), equivalent to the \it actual ideal occurrence \rm of the event \$P_I(\mbox{\bf R})\$ in the final state \$U_{I,II}\rho
_{I,II}(t_i)U^{\dagger }_{I,II}\$
at \$t_f\$. According to the above Lemma on Localization, this occurrence consists in restriction of the state to the region {\bf R} with the corresponding probability.\\

We are interested in the first
subsystem. At the initial moment
\$t_i\$  of the experiment it is,
of course, in the state described
by the reduced density operator:
$$\tr_{II}\Big(Q_{II}\rho _{I,II}(t_i)
Q_{II}\Big)\Big/\tr\Big(
Q_{II}\rho_{I,II}(t_i)\Big),$$
which equals \$\rho_I(t_i)\$ given
by (8b) if one puts
$$\rho _{I,II}(t_i)\equiv\ket{\Phi ,t_i}
_{I,II}\bra{\Phi ,t_i}_{I,II}.$$\\

We make now another short digression.\\

\vspace{0.5cm}

%Ideal occur sec
%%%%%%%%%%%%%%%%%%%%%%%%%%%%%%%%%%%%
\section{Ideal Occurrence}
%%%%%%%%%%%%%%%%%%%%%%%%%%%%%%%%%%%%

Though the concept of {\it ideal occurrence} of an event (projector) \$F\$ in a state (density operator) \$\rho\$, and the corresponding change of state obtained by application of the {\it selective L\"{u}ders formula} $$\rho\quad\rightarrow
\quad F\rho F/\Big(\tr(\rho F)\Big) \eqno{(13)}$$ are known, but perhaps they are not sufficiently well known. Ideal occurrence plays an important role in this study. Therefore, it might be justified to make a few
remarks about it.

In {\it direct interaction} between quantum system and measuring instrument ideal \M of an event (projector) \$F\$ {\it can never occur}. This is obvious from the fact that if the state is an 1-eigenstate of the event (the event has occurred, if it is a property, it is possessed), then \$F\rho = \rho\$ is valid. (This is the algebraic equivalent of the certainty formula \$\tr(\rho F)=1\$. If unfamiliar with this equivalence, see proof in \cite {RAIO2}, Lemma A.4. in Appendix 2 there.) It has the consequence that, as (13) obviously implies, the state should {\it not change at all} in ideal occurrence. This is not possible, because direct interaction in \QM requires exchange of at least one quantum of action. Hence, it must result in change of state.

Ideal occurrence takes place in negative, in distant, and in implied \m . In the above second modification of SG \M we had an example of {\it negative \M } and ideal occurrence (cf subsection 2.3).

An example of {\it distant \M } appeared in the above first modification of the SG \M (cf subsection 2.2). There direct \M was performed on the spatial degree of freedom of the particle, and, by this very act, the spin projection was distantly measured. (The term "distantness", in analogy with two distant but entangled particles, is here used to emphasize that the \M of spin projection is due exclusively to the suitable strong correlations between the spatial and the spin degrees of freedom, which are formally equal to the case of the two material subsystems.)

Finally, ideal occurrence in {\it implied \M } we had in our, perhaps, most important formula (12) in the case of 'being found in the spatial region {\bf R}'.\\

\vspace{0.5cm}

%Evol sec
%%%%%%%%%%%%%%%%%%%%%%%%%%%%%%%%%%%%
\section{Evolution After Preparation}
%%%%%%%%%%%%%%%%%%%%%%%%%%%%%%%%%%%%

Since the evolution operator used
so far applies to the composite
system, there is redundancy in it.
We now eliminate this burden.

Utilizing the identity
\$I_{II}=Q_{II}\enskip +\enskip
Q_{II}^{\perp}\$, we can write
$$U_{I,II}\rho_{I,II}U_{I,II}^{\dag}=
U_{I,II}\Big(Q_{II}+Q_{II}^{\perp}
\Big)\rho_{I,II}U_{I,II}^{\dag}.$$

Further, the screen
and the particle, e. g., do not interact
any longer {\it if the latter has
passed the hole} etc., hence the
evolution operator \$U_{I,II}\$
{\it factorizes tensorically} into the
evolution operator of the particle
\$U_I\$ and that of the screen
\$U_{II}\$. More precisely,
$$U_{I,II}Q_{II}=\Big(U_I\otimes
U_{II}\Big)Q_{II},\eqno{(14)}$$
and, owing to this, we can derive
a simple form of the state of the
particle at \$t_f\$  in the region
{\bf R} (relations (16) and (17)
below).

Some event (corresponding to the
measurement  result) occurs in the region {\bf R} . Since
the measurement  apparatus  is  in
this region, the occurrence of
this event implies, as it was
explained above, the occurrence of
the event \$P_I(\mbox{\bf R})\$. Since this
event is not actually measured
(only implied), we are justified, in accordance with the above Lemma on Localization,
to assume that its occurrence
takes place in the ideal way.
Hence, we take the lhs of (12) as
the relevant composite-system
state, and we replace it by the
rhs of (12) both in the case when
the collapse (occurrence of
\$Q_{II}\$) does take place at
\$t_i\$ and when it is only a
retroactive apparent ideal
occurrence. Then we have
$$\rho_I(t_f)=
\tr_{II}\bigg\{U_{I,II}\bigg[Q_{II}
\rho_{I,II}(t_i)Q_{II}\bigg/\tr\bigg(
Q_{II}\rho_{I,II}(t_i)\bigg)\bigg]
U_{I,II}^{\dagger
}\bigg\}.\eqno{(15)}$$

Taking into account (14), one can
write $$Q_{II}U_{I,II}^{\dag}=
(U_{I,II}Q_{II})^{\dag}=
\Big((U_I\otimes
U_{II})Q_{II}\Big)^{\dag}.$$
Further, let us substitute the
obtained expression in (15) to
obtain $$\rho_I(t_f)=\tr_{II}
\bigg\{\bigg[\Big(U_I\otimes
U_{II}\Big)Q_{II}\bigg]
\rho_{I,II}(t_i)
\bigg[\Big(U_I\otimes
U_{II}\Big)Q_{II}\bigg]^{\dag}
\bigg/\tr\bigg(
Q_{II}\rho_{I,II}(t_i)\bigg)\bigg]
\bigg\}.$$

 Finally, we can take \$U_I\$ and
\$U_I^{\dagger }\$ outside the
partial  trace and we can omit
\$U_{II}\$ and \$U_{II}^{\dagger
}\$ under the partial trace (these
are known partial-trace
identities, which run parallel to
those valid for full traces). Thus
we obtain:
$$\rho_I(t_f)=U_I
\rho _I(t_i)U_I^{\dagger },
\eqno{(16)}$$ where \vspace*{2mm}
$$\rho _I(t_i)\equiv\tr_{II}
\bigg[Q_{II}\rho_{I,II}(t_i)
Q_{II}\bigg/\tr\bigg(Q_{II}\rho
_{I,II}(t_i)\bigg)\bigg].\eqno{(17)}$$\\

\vspace{0.5cm}

%Gen Theor sec
%%%%%%%%%%%%%%%%%%%%%%%%%%%%%%%%%%%%%%
\section{General Theory of Preparation}
%%%%%%%%%%%%%%%%%%%%%%%%%%%%%%%%%%%%%%

To begin with, it should be noted that actual occurrences of events (collapses) are not encompassed by the evolution operator. (This fact is known as the paradox of quantum measurement theory). Therefore, preparation cannot be described by unitary evolution all over. Occurrences of events must be separately included in the change of state in preparation.

In this section we use the same
notation as in the preceding
section. The two subsystems, subsystem \$I\$ on which an experiment is performed and the preparator, subsystem \$II\$,
interact  and reach a {\it
composite-system state} (density
operator) \$\rho _{I,II}(t_i)\$.
This is the first basic
entity of the preparation theory.
The second one is a {\it
triggering event} (projector)
\$Q_{II}\$ on the prepator. The third basic
entity is {\it the prepared
state}, actually the conditional
state \$\rho_I(t_i)\$ of the first
subsystem to which  the occurrence
of the triggering event \$Q_{II}\$
in the composite-system state
\$\rho _{I,II}(t_i)\$ gives rise.
The occurrence may take place in
whatever way, i.e., it need not be
ideal. The conditional state is
given by (17).

Finally, there is an important
fourth entity that belongs
more to the rest of the experiment than
to the preparation. But it is
preparation that must provide the {\it
separate evolution of the object
subsystem}.

In this sense one has the fourth
entity of preparation. It is the
evolution operator
\$U_{I,II}\equiv U_{I,II}(t_f-t_i,t_i)\$
with the important factorization
property (14), which means lack of
interaction between object and
preparator in the interval from
\$t_i\$ till \$t_f\$ after the
triggering event has happened
actually in whatever way or
retroactively apparently in the
ideal way.

As a matter of fact, for a  given
preparator  it  is  only \$U_I\$,
the evolution operator of the
object, that must be known (cf
(14)). As to \$U_{II}\$, the
evolution operator of the
preparator, it is sufficient to
know that it enters the theory via
(14). The concrete form of
\$U_{II}\$ is of no consequence
for the experiment with subsystem
\$I\$.\\

One should note that one assumes
the validity of a unitary
evolution (by
\$U_{I,II}(t_f-t_i,t_i)\$) in the
composite, quantum object plus
preparator, system. This includes
the object-preparator interaction,
but excludes any interaction with
the environment. The unitary
operator part \$U_I(t_f-t_i,t_i)\$ (cf
(14)) of \$U_{I,II}\$ implies that
the object does not interact not
only with the preparator, but also
with the environment. Hence, if
\$\rho_I(t_i)\$ contains coherence
(the interference-creating
property) it will be preserved up
to the final moment in
\$\rho_I(t_f)=U_I\rho_I(t_i)
U^{\dag}_I\$. In other
words, the everywhere lurking {\it
decoherence}, the
coherence-destroying interaction
with the environment, is not
operating because a quantum
experiment is typically
sufficiently dynamically isolated
from the environment.\\

The proposed preparation theory includes only occurrence (any occurrence cf section 4) of events occurring on the preparator. One may wonder what if in some preparation events on the system or on the composite-system (system plus preparator) (or both) occur, and what if no event (actual or fictitious) occurs on the preparator.

Occurrence of possible events on the system or on the composite system in preparation
have to be included in the first above preparation entity, the composite-system state \$\rho_{I,II}(t_i)\$. If no event (actual or fictitious) occurs on the preparator in preparation, then the third preparation entity, the prepared or conditional state is $$\rho_I(t_i)\equiv\tr_{II}\Big(\rho_{I,II} (t_i)\Big)=\tr_{II}\Big(\rho_{I,II}(t_i)
I_{II}\Big).\eqno{(18)}$$ In words, the subsystem state (reduced density operator) is a special case of a conditional state with the condition being the certain event, expressed by the identity operator.\\

\vspace{0.5cm}

%Puzzle sec
%%%%%%%%%%%%%%%%%%%%%%%%%%%%%%%%%%%%%
\section{Puzzling Features}
%%%%%%%%%%%%%%%%%%%%%%%%%%%%%%%%%%%%%

We must distinguish {\it two kinds
of preparators}: the
{\it dynamical} (or immediate-occurrence)
ones, in which the triggering
event does actually occur at
\$t_i\$ (due to a dynamical process, a \m ), and the {\it geometrical} (or delayed-occurrence) ones, in which a special
geometry singles out a spatial
region {\bf R}, and some event (some
measurement result) actually
occurs on the object in {\bf R} at
the delayed moment \$t_f\$. (It is
delayed as far as the preparation
is concerned. It is the final
moment of the experiment.) But, as
explained in detail in section 5, this gives rise
to the retroactive apparent ideal
occurrence (RAIO) of  a
triggering event \$Q_{II}\$, and the entire
theory has exactly the same form
as  for  a dynamical preparator.

One should be aware of some \it
puzzling features \rm of the
preparator theory presented.

(i) We have one and the same
formalism, but two  {\it different
physical mechanisms}, i.e., the
mentioned two kinds of preparators
are equally described, but they
are understood as different
processes.

(ii) The concept of RAIO, which
enables us to describe both kinds
of preparators by the same
formalism, is itself a puzzling
one:

In our hole-in-the-screen example,
intuitively we do feel  that  the
particle must have passed the hole
if it reaches region {\bf R} . But
the conventional \I of \qm , with
the idea of collapse that makes
the events actually occur, seems
to prove us wrong. Since there is
no measurement at \$t_i\$, there
is no collapse and no occurrence
at that moment in actuality.

The composite system is decribed
by
\$\rho_{I,II}(t_f)\enskip\Big[\equiv
U_{I,II}\rho_{I,II}(t_i)U_{I,II}
^{\dag}\Big]\$ at \$t_f\$. This state
may include, possibly in  a
coherent (i.e.,
interference-allowing) way, also
the possibility that the hole is
not passed  in  the described
example. At the final moment
\$t_f\$, and only then, something
happens, some measurement result
is obtained. From the very fact
that  this result is obtained in
the region {\bf R} , we have the
collapse described by the lhs of
(12). It does imply the RAIO of
the triggering event, the rhs of
(12), but this is only formal (or
apparent).

In the standard or conventional
interpretation of \qm , which is
utilized throughout, one does not
search for the mechanism of the
collapse that gives rise to the
occurrence of some event. But,
collapse is taken seriously: it is
considered to be a {\it real,
objectively happening physical
process}. Hence, one is puzzled
that lack of such occurrence,
equally as occurrence, can lead to
correct preparation of an experiment.\\

Besides the embarrassing
appearance of the RAIO in
preparation, there is also the well known
conceptual collision of collapse
with unitary evolution. In
particular, if the latter would
reign by itself, then the prepared
state would always be
\$\tr_{II}\rho_{I,II}\$, and
not
\$\tr_{II}\Big\{Q_{II}\rho_{I,II}Q_{II}
\Big/\Big[\tr\Big(\rho_{I,II}Q_{II}\Big)
\Big]\Big\}\$ (cf (12)) as in the
expounded theory.

At the end of a quantum experiment
one obtains a definite \M result,
though the theory predicts, as a
rule, more than one result. It is
believed to be due to collapse.
This conceptual collision of
collapse and unitary evolution is
well known under the name of 'the
paradox of \Q \M theory'. Since we
have such a conceptual collision
also at the beginning of the
experiment, in preparation, one
should actually speak of {\it the
paradox of the quantum experiment}
with a double collision of the
mentioned basic concepts.

It is important to note that the paradoxes are not due to the proposed formalism; they stem from the fact that there are two kinds of preparations that differ by occurrence and non-occurrence of the triggering event. (Perhaps unfortunately, the proposed theory may seem to sweep this distinction under the carpet.)

The above paradoxes can be
regarded as symptoms of false
interpretation. Many
foundationally-minded physicists
reject the conventional (textbook)
interpretation of \QM precisely
for such reasons.

Since most experiments are performed with ensembles, one sometimes tends to forget that ensembles are many copies of identical individual systems in one and the same individual-system state. (The ensemble and the individual-system state are described by the same density matrix.)

Though ensembles are indispensable because one cannot make practical sense of the probabilities in any other way than as relative frequencies, one cannot understand what is going on in the experiment unless one considers the experiment in the individual-system version as done in this article.\\

\vspace{0.5cm}

%Concl sec
%%%%%%%%%%%%%%%%%%%%%%%%%%%%%%%%%%%%%%%%
\section{Conclusion}
%%%%%%%%%%%%%%%%%%%%%%%%%%%%%%%%%%%%%%%%

It has been shown that, applying the Theorem on RAIO to preparation, a consistently \QMl theory can be obtained.
It is relevant for many experiments, possibly not for all. In this way the first aim (motivation from the Introduction) is achieved.

Also the second aim is achieved. Namely, in the geometrical cases discussed, the RAIO {\it gave full confirmation to the simple, classical intuitive pictorial representation suggested by the geometry of the experiment}. In this lies the answer why preparation has obtained so little attention in the literature: {\it one can lean on classical intuition}.

The goal to treat the dynamical and the geometrical preparations by the same theory has also been achieved.\\

The RAIO is fictitious as far as the standard quantum-mechanical concept of occurrence of an event is concerned. But it is often seen to be quite real {\it concerning classical intuition}. Both in the third example of the Stern-Gerlach and the second example of the hole-in-screen
preparators one has the distinct feeling, based on classical intuition, that the geometry is such that if the particle arrives in the measuring instrument (in whatever way the \M interaction takes place), then it must have passed the preparator in the upper half space or the hole respectively. This makes the RAIO plausible. The point is not plausibility, but the exact \QMl theorem on RAIO (in section 5), which makes possible to describe dynamical and geometrical preparations by the same formalism proposed in this article.\\

Finally, it is important to note that in the proposed theory preparation is almost completely {\it decoupled} from the measurement at the end of the experiment. Actually, the only connection consists in the application of the Theorem on RAIO in the localization formula (12). This is why no mention is made of the kind of \M that is taking place at the end of the experiment.\\

\vspace{0.5cm}

%Ref


\begin{thebibliography}{9}

\bibitem{Bohr}
N. Bohr, {\it Discussion with Einstein on
Epistemological Problems in Atomic
Physics} in {\it Albert Einstein: Philosopher-Scientist},
The Library of Living Philosophers
Inc., editor P. A. Schilpp,  (Evanston, Illinois, 1949), pp. 237-238.

\bibitem{Lud}
G. L\"{u}ders,  Ann. der Physik,
{\bf 8}, 322 (1951).

\bibitem{Messiah}
A. Messiah, {\it Quantum Mechanics},
vol. I, (North Holland, Amsterdam
1961), chapter VIII, subsection I.2.

\bibitem{Laloe}
C. Cohen-Tannoudji, B. Diu and F.
Lalo\'{e}, {\it Quantum Mechanics},
Volume I, (John Wiley and Sons, New York
1977), chapter III, section C.

\bibitem{FHMV'76}
F. Herbut and M. Vuji\v{c}i\'{c},
Ann. Phys. (N. Y.), {\bf 96}, 382
(1976).

\bibitem{FHMV'84}
M. Vuji\v{c}i\'{c} and F. Herbut,
J. Math. Phys., {\bf 25}, 2253
(1984).

\bibitem{FHPhys.Rev.}
F. Herbut, Phys. Rev. A, {\bf 66},
052321 (2002).

\bibitem{Gleason}
A. M. Gleason, J. Math. Mech., {\bf 6},
885 (1957).

\bibitem{RAIO2}
F. Herbut, Found. Phys. Lett.,
{\bf 9}, 437 (1996).




\end{thebibliography}
\end{document}